\documentclass[aps,onecolumn,letterpaper,oneside,preprint,tightenlines,draft,%
nobibnotes,nofootinbib,amsfonts,amssymb,amsmath,eqsecnum,showpacs,%
showkeys]{revtex4}

\newcommand{\cK}{\mathcal{K}}
\newcommand{\cN}{\mathcal{N}}
\newcommand{\bbN}{\mathbb{N}}
\newcommand{\bbZ}{\mathbb{Z}}
\newcommand{\bbR}{\mathbb{R}}

\newcommand{\rmd}{\mathrm{d}}
\newcommand{\rmL}{\mathrm{L}}
\newcommand{\rmT}{\mathrm{T}}
\newcommand{\del}{\partial}
\newcommand{\pprime}{\prime\prime}
\newcommand{\ppprime}{\prime\prime\prime}

\begin{document}


%
%

\title{On General Form of $\cN$-fold Supersymmetry}
\author{Toshiaki Tanaka}
\email{toshiaki@post.kek.jp}
\affiliation{Institute of Particle and Nuclear Studies,
 High Energy Accelerator Research Organization (KEK),
 1-1 Oho, Tsukuba, Ibaraki 305-0801, Japan}


\begin{abstract}

We analyze general structure of $\cN$-fold supersymmetry which provides
a systematic framework to construct weakly quasi-solvable quantum mechanical
systems. Main ingredients of our analysis are dimensional analysis and introduction
of an equivalent class of linear differential operators associated with $\cN$-fold
supersymmetry for each $\cN$. To illustrate how they work, we construct the
most general form of $\cN$-fold supersymmetric systems for $\cN=2$, $3$, and
$4$.

\end{abstract}





\preprint{TH-1472}

\maketitle

\section{Introduction}
\label{sec:intro}

$\cN$-fold supersymmetry (SUSY)~\cite{AIS93,AST01b,AS03} is one of
the most powerful frameworks for constructing a one-dimensional quantum
mechanical (QM) system which admits analytic solutions in closed form in
a certain sense. This is due to the fact that $\cN$-fold SUSY is essentially
equivalent to weak quasi-solvability which is until now the least restrictive
concept about the availability of solutions in closed form. The latter crucial
fact was first proved in a general fashion in Ref.~\cite{AST01b} and was later
complemented slightly in Ref.~\cite{Ta03a}. For a review, see, e.g., Ref.~\cite{Ta09}.

In general, construction of an $\cN$-fold SUSY system get more difficult as
the number $\cN\in\bbN$ increases since we must solve coupled nonlinear
differential equations for $\cN$ unknown functions (see the next section).
To bypass the latter difficulty, a systematic algorithm for constructing an
$\cN$-fold SUSY system based on quasi-solvability (in the strong sense) was
proposed in Ref.~\cite{GT05}. A key ingredient of the algorithm is to choose
first an $\cN$-dimensional linear space of specific functions such that it can be
preserved by a second-order linear differential operator. It has been proved
to be quite efficient and so far four inequivalent types of $\cN$-fold SUSY,
namely, type A~\cite{AST01a,Ta03a}, type B~\cite{GT04}, type C~\cite{GT05},
and type $X_{2}$~\cite{Ta10a} are successfully constructed with the algorithm.
We note that almost all the models having essentially the same symmetry as
$\cN$-fold SUSY but called with other terminologies in the literature, such as
P\"{o}schl--Teller and Lam\'{e} potentials, are actually particular cases of type A
$\cN$-fold SUSY.

It is evident, however, that the algorithm is helpless to construct a \emph{weakly}
quasi-solvable system which only admits a finite-dimensional invariant subspace
determined by another differential equation. The framework of $\cN$-fold SUSY
covers such systems and thus provides a more general formalism than
higher-derivative generalizations of Darboux transformation such as the Crum's
method~\cite{Cr55a} which relies on a set of exact eigenfunctions of a regular 
Sturm--Liouville system. To construct a weakly quasi-solvable system, we must
in general treat directly the aforementioned coupled nonlinear differential equations.
For the simplest case of $\cN=2$, the general result was already studied and
reported in Refs.~\cite{AICD95,AIN95b,AST01b}. On the other hand, for the cases
of $\cN>2$ until now on there is, at the best of our knowledge, only one
paper~\cite{IN04} which studied the $\cN=3$ case. This fact would reflect
the difficulty and complexity of the problems for larger $\cN$.

In this work, we investigate general strucuture of $\cN$-fold SUSY systems
to extract relevant clues to construct them which have in particular weak
quasi-solvability. For this purpose, we first employ dimensional
analysis which is a well-known powerful tool in general physics. It turns out
that it is also quite efficient in acquiring deeper understandings of $\cN$-fold
SUSY. We then introduce equivalent classes of linear differential operators
associated with $\cN$-fold SUSY. We find that it enables us to deal with
operator equalities appeared in $\cN$-fold SUSY more systematically and
transparently.

We organize the paper as follows. In the next section, we first briefly review
the ingredients of $\cN$-fold SUSY. Then, we introduce two key concepts
for analyzing its general structure, namely, dimensional analysis and equivalent
classes of linear differential operators associated with $\cN$-fold SUSY.
In Sections~\ref{sec:2f}--\ref{sec:3f}, we apply the general arguments to obtain
general form of $\cN$-fold SUSY for $\cN=2$, $3$, and $4$, respectively.
We show how dimensional analysis enables us to reduce the complexity of
the problems on solving the conditions for $\cN$-fold SUSY and on finding
integral constants of the systems. In the last section, we summarize the paper
and provide comments on the future issues.

\section{General Consideration}
\label{sec:gcon}

To begin with, we shall briefly review ingredients of $\cN$-fold SUSY as
preliminaries. For details, see the review~\cite{Ta09}. An $\cN$-fold SUSY
QM system in one-dimension is composed of a pair of Hamiltonians $H^{\pm}$
and a pair of $\cN$th-order linear differential operators $P_{\cN}^{\pm}$
\begin{align}
H^{\pm}=-\frac{1}{2}\frac{\rmd^{2}}{\rmd q^{2}}+V^{\pm}(q),\quad
 P_{\cN}^{-}=\frac{\rmd^{\cN}}{\rmd q^{\cN}}+\sum_{k=0}^{\cN-1}
 w_{k}^{[\cN]}(q)\frac{\rmd^{k}}{\rmd q^{k}},\quad
 P_{\cN}^{+}=(P_{\cN}^{-})^{\rmT},
\label{eq:HPN}
\end{align}
where the superscript T denotes the transposition of a linear operator
\cite{AS03}, which satisfy the intertwining relation
\begin{align}
P_{\cN}^{-}H^{-}-H^{+}P_{\cN}^{-}=\sum_{k=0}^{\cN}I_{k}^{[\cN]}
 \frac{\rmd^{k}}{\rmd q^{k}}=0,
\label{eq:inter}
\end{align}
and its transposed relation $H^{-}P_{\cN}^{+}-P_{\cN}^{+}H^{+}=0$. The operators
$P_{\cN}^{\pm}$ are actually components of $\cN$-fold supercharges.

One of the most significant consequences of the intertwining relation
(\ref{eq:inter}) is \emph{weak quasi-solvability}~\cite{AST01b,Ta03a}.
That is, each $\cN$-fold SUSY Hamiltonian $H^{\pm}$ preserves the linear
space $\ker P_{\cN}^{\pm}$:
\begin{align}
H^{\pm}\ker P_{\cN}^{\pm}\subset\ker P_{\cN}^{\pm}.
\end{align}
If the differential equation $P_{\cN}^{-}\phi=0$ and/or $P_{\cN}^{+}\phi=0$
admits a number of analytic solutions in closed form, $H^{-}$ and/or $H^{+}$
is not only weakly quasi-solvable but also \emph{quasi-solvable} in the strong
sense. But in general, an $\cN$-fold SUSY Hamiltonian is merely weakly
quasi-solvable and does not admit any analytic local solutions. We also note that
$\ker P_{\cN}^{\pm}$ is not necessarily a subspace of the linear space,
which is usually the Hilbert space $L^{2}(S)$ ($S\subset\bbR$), in which
the operator $H^{\pm}$ acts.

Another peculiar feature of an $\cN$-fold SUSY system is that the product
$P_{\cN}^{\mp}P_{\cN}^{\pm}$ which arises as a component of the anti-commutator
of $\cN$-fold supercharges is an $\cN$th-degree polynomial in the Hamiltonian
$H^{\pm}$~\cite{AST01b, AS03} and thus has the following form:
\begin{align}
P_{\cN}^{\mp}P_{\cN}^{\pm}=2^{\cN}\left[(H^{\pm}+C_{0})^{\cN}
 +\sum_{k=1}^{\cN-1}C_{k}(H^{\pm}+C_{0})^{\cN-k-1}\right],
\label{eq:PNPN}
\end{align}
where $C_{k}$ ($k=0,\dots,\cN-1$) are all constants. The $\cN$ zeros of
the polynomial in the r.h.s.\ of (\ref{eq:PNPN}) correspond to the spectrum of
$H^{\pm}$ in the space $\ker P_{\cN}^{\pm}$. Hence, they are actually
a part of the eigenvalues of $H^{\pm}$ if $\ker P_{\cN}^{\pm}\subset L^{2}(S)$.
In the latter case, we can calculate the part of the eigenvalues algebraically
from (\ref{eq:PNPN}) even though the corresponding eigenfunctions cannot
be obtained in closed form. We note that the $\cN$ energy spectra would
exhibit the characteristic features such as the disappearance of nonperturbative
corrections due to the generalized non-renormalization theorem in $\cN$-fold
SUSY, which were reported on the realistic physical systems such as the asymmetric
double-well potential in \cite{AKOSW99} and the periodic and the symmetric
triple-well potentials in \cite{ST02}.

One of the most difficult problems on $\cN$-fold SUSY is to analyze the
condition (\ref{eq:inter}) for $\cN$-fold SUSY. It is composed of coupled
nonlinear differential equations for the so far undetermined functions
$w_{k}^{[\cN]}$ ($k=0,\dots,\cN-1$). As is easily expected, its complexity
gets terrible as the integer $\cN$ increases. Hence, it is quite difficult to
solve directly the condition (\ref{eq:inter}) for larger $\cN$. In the subsequent
two sections, we shall investigate general aspects of the complicated
structure of $\cN$-fold SUSY systems which would provide us a clearer view
on them.

\subsection{Dimensional Analysis}
\label{ssec:da}

Dimensional analysis is one of the powerful methods to make physical consideration
and in particular to estimate a physical quantity under consideration without solving
equations directly, cf., any textbook on general physics. In this section, we shall
see that it supplies us with a valuable guiding principle in solving the condition for
$\cN$-fold SUSY.

To make dimensional analysis on our system (\ref{eq:HPN}), we first note that
we have implicitly employed the unit system where the ratio of square action
(the Planck constant) to mass $\hbar^{2}/m$ is dimensionless. The only relevant
physical dimension is then the \emph{length}, denoted by $[\rmL]$, which is carried
by the physical position variable $q$. It is easy to see from (\ref{eq:HPN}) and
(\ref{eq:PNPN}) that the physical dimensions of $V^{\pm}$, $C_{k}$, and $w_{k}^{[\cN]}$
in terms of the length are given by
\begin{align}
\begin{split}
&V^{\pm}\ [\rmL^{-2}],\qquad C_{k}\ [\rmL^{-2(k+1)}]\quad (k=0,\dots,\cN-1),\\
&w_{k}^{[\cN](m)}\ [\rmL^{k-\cN-m}]\quad(k=0,\dots,\cN-1,\, m=0,1,2,\ldots),
\label{eq:pdim1}
\end{split}
\end{align}
where $w_{k}^{[\cN](m)}(q)$ is the $m$th derivative of $w_{k}^{[\cN]}(q)$ with
respect to $q$.

Dimensional analysis relies on the obvious fact that all the terms which appear
in a single formula under consideration must have the same physical dimension.
For instance, a potential has the physical dimension $[\rmL^{-2}]$ and thus
must be expressed as a sum of terms all of which have the same physical
dimension $[\rmL^{-2}]$. Hence, if we only consider a polynomial of $C_{k}$
and $w_{k}^{[\cN](m)}$ ($m=0,1,2,\ldots$), it must have the following form
\begin{align}
V=\alpha_{0}w_{\cN-2}^{[\cN]}+\alpha_{1}w_{\cN-1}^{[\cN]\prime}
 +\alpha_{2}\bigl(w_{\cN-1}^{[\cN]}\bigr)^{2}-C_{0}\ [\rmL^{-2}],
\label{eq:Vform}
\end{align}
where $\alpha_{k}$ ($k=0,1,2$) are all dimensionless parameters. In fact, we can
see that a pair of $\cN$-fold SUSY potentials $V^{\pm}$ satisfying (\ref{eq:inter})
does have the form (\ref{eq:Vform}). The l.h.s.\ of (\ref{eq:inter}) is a linear
differential operator of at most $\cN$th order, as is indicated in (\ref{eq:inter}),
and it is evident that the identity (\ref{eq:inter}) holds if and only if all the coefficients
$I_{k}^{[\cN]}$ of $\del^{k}=\rmd^{k}/\rmd q^{k}$ ($k=0,\dots,\cN$) vanish. The latter
requirement for the coefficients of $\del^{\cN}$ and $\del^{\cN-1}$ reads as
\begin{subequations}
\label{eqs:Nfc1}
\begin{align}
I_{\cN}^{[\cN]}&=w_{\cN-1}^{[\cN]\prime}-(V^{+}-V^{-})=0\ [\rmL^{-2}],\\
2I_{\cN-1}^{[\cN]}&=w_{\cN-1}^{[\cN]\pprime}+2w_{\cN-2}^{[\cN]\prime}
 +2\cN V^{-\prime}-2w_{\cN-1}^{[\cN]}(V^{+}-V^{-})=0\ [\rmL^{-3}].
\end{align}
\end{subequations}
The set of conditions (\ref{eqs:Nfc1}) can be easily solved as
\begin{align}
V^{\pm}=-\frac{1}{\cN}w_{\cN-2}^{[\cN]}+\left(\frac{\cN-1}{2\cN}\pm
 \frac{1}{2}\right)w_{\cN-1}^{[\cN]\prime}+\frac{1}{2\cN}\bigl(w_{\cN-1}^{[\cN]}
 \bigr)^{2}-C_{0}\ [\rmL^{-2}],
\label{eq:NfgV}
\end{align}
which indeed has the form of (\ref{eq:Vform}). The formula (\ref{eq:NfgV})
provides a general expression for a pair of $\cN$-fold SUSY potentials $V^{\pm}$
for an arbitrary $\cN\in\bbN$.
One of its characteristic features is that they are expressible solely in terms of
the two functions $w_{\cN-1}^{[\cN]}$ and $w_{\cN-2}^{[\cN]}$ irrespective of
what additional conditions $w_{k}^{[\cN]}$ ($k=0,\dots,\cN-1$) should satisfy.

The remaining conditions for $\cN$-fold SUSY coming from the coefficients
of $\del^{k}$ for $k=0,\dots,\cN-2$ in (\ref{eq:inter}) are in general algebraic
equations consisting of $w_{k}^{[\cN](m)}$ ($m=0,1,2,\ldots$) after
the substitution of (\ref{eq:NfgV}) into (\ref{eq:inter}).
Dimensional analysis tells us that the operator in the l.h.s.\ of (\ref{eq:inter})
has the physical dimension $[\rmL^{-\cN-2}]$ and thus $I_{k}^{[\cN]}$
($k=0,\dots,\cN-2$) has the physical dimension $[\rmL^{k-\cN-2}]$.
For instance, $I_{\cN-2}^{[\cN]}$ and $I_{\cN-3}^{[\cN]}$ are calculated as
\begin{align}
-4\cN I_{\cN-2}^{[\cN]}=&\;\cN(\cN-1)w_{\cN-1}^{[\cN]\ppprime}+2\cN(\cN-2)
 w_{\cN-2}^{[\cN]\pprime}-4\cN w_{\cN-3}^{[\cN]\prime}-2(\cN-1)^{2}
 w_{\cN-1}^{[\cN]}w_{\cN-1}^{[\cN]\pprime}\notag\\
&\;-2\cN(\cN-1)\bigl(w_{\cN-1}^{[\cN]\prime}\bigr)^{2}
 +4\cN w_{\cN-1}^{[\cN]\prime}
 w_{\cN-2}^{[\cN]}+4(\cN-1)w_{\cN-1}^{[\cN]}w_{\cN-2}^{[\cN]\prime}\notag\\
&\;-4(\cN-1)\bigl(w_{\cN-1}^{[\cN]}\bigr)^{2}w_{\cN-1}^{[\cN]\prime}\ [\rmL^{-4}],
\end{align}
\begin{align}
\lefteqn{
-12\cN I_{\cN-3}^{[\cN]}=\cN(\cN-1)(\cN-2)\bigl(w_{\cN-1}^{[\cN]
 \pprime\pprime}
 +2w_{\cN-2}^{[\cN]\ppprime}\bigr)-6\cN\bigl(w_{\cN-3}^{[\cN]\pprime}
 +2w_{\cN-4}^{[\cN]\prime}\bigr)
}\notag\hspace{30pt}\\
&\;-(\cN-1)(\cN-2)\Bigl[(2\cN-3)w_{\cN-1}^{[\cN]}w_{\cN-1}^{[\cN]\ppprime}
 +6\cN w_{\cN-1}^{[\cN]\prime}w_{\cN-1}^{[\cN]\pprime}\Bigr]\notag\\
&\;+6(\cN-2)\Bigl[w_{\cN-1}^{[\cN]\pprime}w_{\cN-2}^{[\cN]}
 +(\cN-1)w_{\cN-1}^{[\cN]}w_{\cN-2}^{[\cN]\pprime}\Bigr]+12\cN
 w_{\cN-1}^{[\cN]\prime}w_{\cN-3}^{[\cN]}\notag\\
&\;+12(\cN-2)w_{\cN-2}^{[\cN]}
 w_{\cN-2}^{[\cN]\prime}-6(\cN-1)(\cN-2)\Bigl[\bigl(w_{\cN-1}^{[\cN]}\bigr)^{2}
 w_{\cN-1}^{[\cN]\pprime}+w_{\cN-1}^{[\cN]}\bigl(w_{\cN-1}^{[\cN]\prime}\bigr)^{2}
 \Bigr]\notag\\
&\;-12(\cN-2)w_{\cN-1}^{[\cN]}w_{\cN-1}^{[\cN]\prime}w_{\cN-2}^{[\cN]}
 \ [\rmL^{-5}],
\end{align}
and consist of the terms which are consistent with the dimensional analysis.

Ideally, one can obtain a general form of $\cN$-fold SUSY systems if one
succeeds in expressing all the $\cN$ functions $w_{k}^{[\cN]}$ ($k=0,\dots,\cN-1$),
which characterize the systems, in terms of a single function, say, $u$ and its
derivatives $u'$, $u''$, $\ldots$ by solving the set of the $\cN-1$ constraints
$I_{k}^{[\cN]}=0$ ($k=0,\dots,\cN-1$). If it is eventually the case, we have a set
of $\cN$ functionals $u_{k}^{[\cN]}$ ($k=0,\dots,\cN-1$) such that
\begin{align}
w_{k}^{[\cN]}=u_{k}^{[\cN]}[u]\ [\rmL^{k-\cN}]\quad(k=0,\dots,\cN-1).
\label{eq:defuk}
\end{align}
One of the most important aspects of (\ref{eq:defuk}) is that each $u_{k}^{[\cN]}$
($k=0,\dots,\cN-1$) has the same physical dimension as the one of $w_{k}^{[\cN]}$.
It in particular means that there would be a set of transformations $w_{k}^{[\cN]}
\to u_{k}^{[\cN]}$ which preserve all the physical dimensions. Conversely, if we
can find a set of dimension-preserving transformations $w_{k}^{[\cN]}\to
u_{k}^{[\cN]}$, we may solve the set of the constraints more easily. In
Sections~\ref{sec:2f}--\ref{sec:4f}, we will employ this strategy to see how
drastically we can reduce the complexity of the constraints.

To solve the constraints to get (\ref{eq:defuk}) is in principle possible unless some
of the constraints automatically imply others since we have $\cN-1$ constraints
for the $\cN$ unknown functions. However, the task would get drastically harder
as the integer $\cN$ increases.
One of the clues to circumvent the situation is in (\ref{eq:PNPN}). All the
coefficients of derivative operators (except for the highest $2\cN$th-order
and including the lowest $0$th-order) in the l.h.s.\ of (\ref{eq:PNPN}) are
quadratic forms of $w_{k}^{[\cN](m)}$ while the r.h.s.\ depends on, in addition
to the potentials $V^{\pm}$, the $\cN$ constants $C_{k}$ ($k=0,\dots,\cN-1$)
which are absent in the l.h.s. The latter fact indicates the existence of $\cN$
integral constants of any $\cN$-fold SUSY system which would be functionals
of $w_{j}^{[\cN]}$ ($j=0,\dots,\cN-1$) and $V^{\pm}$ whose physical dimensions
are the same as the ones of $C_{k}$:
\begin{align}
C_{k}=J_{k}[w^{[\cN]}, V]\ [\rmL^{-2(k+1)}]\quad(k=0,\dots,\cN-1).
\end{align}
They must emerge from the integration of the set of differential equations
\begin{align}
\frac{\rmd}{\rmd q}J_{k}[w^{[\cN]}, V]=0\ [\rmL^{-(2k+3)}]\quad(k=0,\dots,\cN-1).
\label{eq:dJdq}
\end{align}
The latter equations are another set of constraints. On the other hand,
the set of equalities $I_{k}^{[\cN]}=0$ ($k=0,\dots,\cN$) are the only
constraints which come from the condition for $\cN$-fold SUSY. Hence,
the differential equations (\ref{eq:dJdq}) must be such equations that hold
whenever all the conditions $I_{k}^{[\cN]}=0$ ($k=0,\dots,\cN$) are
satisfied. This means that all the quantities $\rmd J_{k}/\rmd q$
($k=0,\dots,\cN-1$) in the l.h.s.\ of (\ref{eq:dJdq}) would be expressible in terms
of $I_{j}^{[\cN]}$ in a way such that the identities $I_{j}^{[\cN]}=0$
($j=0,\dots,\cN$) apparently imply (\ref{eq:dJdq}). The most general form
of such a kind would be
\begin{align}
\frac{\rmd J_{k}}{\rmd q}=\sum_{j=0}^{\cN}L_{kj}I_{j}^{[\cN]}\ [\rmL^{-(2k+3)}]
 \quad(k=0,\dots,\cN-1),
\label{eq:dJdq1}
\end{align}
where $L_{kj}$ are all linear differential operators whose coefficients consist
of only $w_{k}^{[\cN]}$, $V^{\pm}$, and their derivatives.
Each $L_{kj}$ must has the physical dimension $[L^{\cN-2k-j-1}]$
since the ones of $\rmd J_{k}/\rmd q$ and $I_{j}^{[\cN]}$ are $[\rmL^{-(2k+3)}]$
and $[\rmL^{\cN-2-j}]$, respectively.

To see the validity of the above argument, let us consider the constant $C_{0}$.
{}From (\ref{eq:NfgV}), we immediately know the form of $J_{0}$ as
\begin{align}
C_{0}=J_{0}[w^{[\cN]}, V]=-V^{-}-\frac{1}{\cN}w_{\cN-2}^{[\cN]}+\frac{1}{2\cN}
 w_{\cN-1}^{[\cN]\prime}+\frac{1}{2\cN}\bigl(w_{\cN-1}^{[\cN]}\bigr)^{2}
 \ [\rmL^{-2}].
\end{align}
Hence, the differential equation which leads to the latter equation is
\begin{align}
0=\frac{\rmd J_{0}}{\rmd q}=-V^{-\prime}-\frac{1}{\cN}w_{\cN-2}^{[\cN]\prime}
 +\frac{1}{2\cN}w_{\cN-1}^{[\cN]\pprime}+\frac{1}{\cN}w_{\cN-1}^{[\cN]}
 w_{\cN-1}^{[\cN]\prime}\ [\rmL^{-3}].
\label{eq:J0}
\end{align}
We then check by using (\ref{eqs:Nfc1}) that the r.h.s.\ of (\ref{eq:J0}) can
be in fact expressed in terms of $I_{j}^{[\cN]}$ as
\begin{align}
\frac{\rmd J_{0}}{\rmd q}=\frac{1}{\cN}w_{\cN-1}^{[\cN]}I_{\cN}^{[\cN]}
 -\frac{1}{\cN}I_{\cN-1}^{[\cN]}\ [\rmL^{-3}],
\end{align}
which indeed has the form of (\ref{eq:dJdq1}) with
\begin{align}
L_{0\cN}=\frac{1}{\cN}w_{\cN-1}^{[\cN]}\ [\rmL^{-1}],\quad
 L_{0\,\cN-1}=-\frac{1}{\cN}\ [\rmL^{0}],\quad L_{0k}=0\ (k=0,\dots,\cN-2),
\end{align}
all having the correct physical dimensions $[\rmL^{\cN-j-1}]$ for $L_{0j}$
($j=0,\dots,\cN$).

In practice, we already solved the two conditions (\ref{eqs:Nfc1}) to obtain
the general form of $V^{\pm}$ as (\ref{eq:NfgV}), and thus we can totally
eliminate $V^{\pm}$ in the remaining conditions $I_{k}^{[\cN]}=0$ ($k=0,\dots,
\cN-2$). As a result, the remaining integral constants $C_{k}$ ($k=1,\dots,\cN-1$)
would be functionals of only $w_{j}^{[\cN]}$ ($j=0,\dots,\cN-1$):
\begin{align}
C_{k}=J_{k}[w^{[\cN]},V[w^{[\cN]}]]:=J_{k}[w^{[\cN]}]\ [\rmL^{-2(k+1)}]
 \quad(k=1,\dots,\cN-1).
\end{align}
Accordingly, $\rmd J_{k}/\rmd q$ ($k=1,\dots,\cN-1$) would be expressed
in terms of the remaining $I_{j}^{[\cN]}$ ($j=0,\dots,\cN-2$) as
\begin{align}
0=\frac{\rmd J_{k}}{\rmd q}=\sum_{j=0}^{\cN-2}L_{kj}I_{j}^{[\cN]}\ [\rmL^{-(2k+3)}]
 \quad(k=1,\dots,\cN-1).
\label{eq:dJdq2}
\end{align}
We will later see in Sections~\ref{sec:2f}--\ref{sec:4f} that the above analysis
is actually valid and helps us to obtain general forms of $\cN$-fold SUSY for
$\cN=2$, $3$, and $4$.

To summarize, the existence of $\cN$ constants $C_{k}$ ($k=0,\dots,\cN-1$)
in the r.h.s. of (\ref{eq:PNPN}) has the direct relation to the existence of $\cN$
constraints $I_{k}^{[\cN]}=0$ ($k=0,\dots,\cN-1$) which are differential equations
after eliminating the algebraic constraint $I_{\cN}^{[\cN]}=0$. The physical
dimensions of the relevant quantities are
\begin{align}
I_{k}^{[\cN]}\ [\rmL^{k-\cN-2}],\quad u_{k}^{[\cN]}\ [\rmL^{k-\cN}],\quad
 J_{k}\ [\rmL^{-2(k+1)}],\quad L_{kj}\ [\rmL^{\cN-2k-j-1}].
\end{align}

\subsection{Equivalent Classes of Linear Differential Operators}
\label{ssec:eqc}

The existence of $\cN-1$ constraints $I_{j}^{[\cN]}=0$ ($j=0,\dots,\cN-2$)
after the determination of the potential pair (\ref{eq:NfgV}) results in the
existence of null operators associated with the $\cN$-fold SUSY system under
consideration. Let us first introduce a linear space of linear differential operators,
denoted by $\cK^{[\cN]}$, whose coefficients are all functionals of only
$w_{k}^{[\cN]}$ ($k=0,\dots,\cN-1$). Let $K_{ij}[w^{[\cN]}]\in\cK^{[\cN]}$
($i=0,1,2,\ldots$) and define a set of functionals $f_{i}^{[\cN]}[w^{[\cN]}]$ by
\begin{align}
f_{i}^{[\cN]}[w^{[\cN]}]=\sum_{j=0}^{\cN-2}K_{ij}[w^{[\cN]}]I_{j}^{[\cN]}.
\label{eq:deffi}
\end{align}
We then define a subspace of $\cK^{[\cN]}$, denoted by $\cK_{0}^{[\cN]}$,
which consists of linear differential operators whose coefficients are all given
by $f_{i}^{[\cN]}$ introduced in (\ref{eq:deffi}). That is, $K_{0}\in\cK_{0}^{[\cN]}$
means that there exists a set of linear differential operators $K_{ij}[w^{[\cN]}]\in
\cK^{[\cN]}$ such that
\begin{align}
K_{0}=\sum_{i}f_{i}^{[\cN]}[w^{[\cN]}]\del_{i}=\sum_{i}\left(\sum_{j=0}^{\cN-2}
 K_{ij}[w^{[\cN]}]I_{j}^{[\cN]}\right)\del_{i}.
\label{eq:K0}
\end{align}
It is obvious by definition that any element of $\cK_{0}^{[\cN]}$ is a null operator
so long as all the $\cN$-fold SUSY constraints $I_{j}^{[\cN]}=0$ ($j=0,\dots,\cN-2$)
are satisfied. Hence, the linear space in which an $\cN$-fold SUSY system is
considered is actually the quotient space $\cK^{[\cN]}/\cK_{0}^{[\cN]}$. It naturally
leads us to introduce an equivalence class of linear differential oerators in
$\cK^{[\cN]}$.
We shall say that two linear differential operators $L_{1}, L_{2}\in\cK^{[\cN]}$
belong to \textit{equivalent class associated with $\cN$-fold supersymmetry}
and express the equivalence as $L_{2}\stackrel{\cN}{\sim}L_{1}$ if $L_{2}-L_{1}
\in\cK_{0}^{[\cN]}$. Any equality between operators $L_{2}=L_{1}$ appeared in
an $\cN$-fold SUSY system should be thus regarded as an equivalent relation
$L_{2}\stackrel{\cN}{\sim}L_{1}$ of the latter equivalent class. In particular,
any explicit expression for a specific operator such as $H^{\pm}$ and
$P_{\cN}^{\pm}$ should be considered as a representative of it with respect
to the equivalent class. In what follows, we will employ the equivalence relation
$\stackrel{\cN}{\sim}$ only when we would like to stress that the left and the
right hand sides of the formula under consideration is identical if and only if
(some of) the constraints $I_{j}^{[\cN]}=0$ ($j=0,\cdots,\cN-2$) are satisfied.

\section{2-fold Supersymmetry}
\label{sec:2f}

It is quite instructive to see how the general consideration in the previous section
make sense in the case of $2$-fold SUSY though its general form was already
obtained by the direct integrations of the constraints~\cite{AICD95,AIN95b,AST01b}.
Components of $2$-fold supercharges are given by
\begin{align}
P_{2}^{-}=\del^{2}+w_{1}\del+w_{0},\qquad
 P_{2}^{+}=\del^{2}-w_{1}\del+w_{0}-w'_{1},
\label{eq:2fP}
\end{align}
where and hereafter we shall omit the superscript $[\cN]$ of $w_{k}^{[\cN]}$
etc.\ for the simplicity unless the omission would not cause any ambiguity or
confusion. The condition for $2$-fold SUSY $P_{2}^{-}H^{-}-H^{+}P_{2}^{-}=0$ is
satisified if and only if the following three equalities hold:
\begin{align}
&V^{+}-V^{-}=w'_{1},
\label{eq:2fc1}\\
&w''_{1}+2w'_{0}+4V^{-\prime}-2w_{1}(V^{+}-V^{-})=0,
\label{eq:2fc2}\\
&w''_{0}+2V^{-\pprime}+2w_{1}V^{-\prime}-2w_{0}(V^{+}-V^{-})=0.
\label{eq:2fc3}
\end{align}
Substituting (\ref{eq:2fc1}) into (\ref{eq:2fc2}) and (\ref{eq:2fc3}), and
integrating the resulting equation from (\ref{eq:2fc2}), we obtain
\begin{align}
4V^{+}&=3w'_{1}-2w_{0}+(w_{1})^{2}-4C_{0},\\
4V^{-}&=-w'_{1}-2w_{0}+(w_{1})^{2}-4C_{0},\\
-4I_{0}&=w'''_{1}-w_{1}w''_{1}-2(w'_{1})^{2}+4w'_{1}w_{0}+2w_{1}w'_{0}-2(w_{1})^{2}w'_{1}=0,
\label{eq:2fc3'}
\end{align}
where $C_{0}$ is an integral constant. To integrate the third equation (\ref{eq:2fc3'}),
we shall first make a dimension-preserving transformation $w_{0}\to u_{0}$. We choose
it such that it will convert simultaneously both the pairs $P_{2}^{\pm}$ and $V^{\pm}$
into symmetric forms. The most general transformation of polynomial type preserving
the physical dimension $[\rmL^{-2}]$ of $w_{0}$ would be
\begin{align}
w_{0}=u_{0}+\frac{1}{2}w'_{1}-\alpha_{0}(w_{1})^{2}\ [\rmL^{-2}],
\label{eq:2ftf}
\end{align}
where $u_{0}\ [\rmL^{-2}]$ and $\alpha_{0}$ is a dimensionless parameter.
We note that the latter transformation indeed renders both the pair of $2$-fold
supercharge components and the pair of potentials of symmetric forms as
\begin{align}
P_{2}^{\pm}&=\del^{2}\mp w_{1}\del+u_{0}-\alpha_{0}(w_{1})^{2}\mp\frac{1}{2}w'_{1},
\label{eq:P2+-}\\
4V^{\pm}&=-2u_{0}+(2\alpha_{0}+1)(w_{1})^{2}\pm2w'_{1}-4C_{0}.
\label{eq:V2+-}
\end{align}
With the transformation (\ref{eq:2ftf}), the condition (\ref{eq:2fc3'}) reads as
\begin{align}
-4I_{0}=w'''_{1}+4w'_{1}u_{0}+2w_{1}u'_{0}-2(4\alpha_{0}+1)(w_{1})^{2}w'_{1}=0
\ [\rmL^{-4}].
\label{eq:2fc3''}
\end{align}
Hence, it gets simplest when
\begin{align}
\alpha_{0}=-1/4.
\label{eq:a0}
\end{align}
The next task we should do is to construct the total differential $\rmd J_{1}/\rmd q$
in (\ref{eq:dJdq2}). We note that $\rmd J_{1}/\rmd q$ and $I_{0}$ have the physical
dimensions $[\rmL^{-5}]$ and $[\rmL^{-4}]$, respectively. Hence, the operator
$L_{10}$ defined in (\ref{eq:dJdq2}) in this case must have the physical dimension 
$[\rmL^{-1}]$. Except for the differential operator $\rmd/\rmd q$, there is
essentially only one multiplicative operator of polynomial type which have that
dimension, namely, $L_{10}\propto w_{1}\ [\rmL^{-1}]$. In fact, we can easily check
that $w_{1}I_{0}$ is of a total differential form and thus we put
\begin{align}
16\frac{\rmd J_{1}}{\rmd q}=-8w_{1}I_{0}=2w_{1}w'''_{1}+8w_{1}w'_{1}u_{0}
 +4(w_{1})^{2}u'_{0}=0\ [\rmL^{-5}].
\end{align}
The latter differential equation is integrated to yield
\begin{align}
16J_{1}[w]=2w_{1}w''_{1}-(w'_{1})^{2}+4(w_{1})^{2}u_{0}=16C_{1}\ [\rmL^{-4}],
\label{eq:2fC1}
\end{align}
where $C_{1}$ is another integral constant having the correct physical dimension
$[\rmL^{-4}]$ listed in (\ref{eq:pdim1}). Hence, we can express $u_{0}$,
and thus $w_{0}$ as well, in terms of $w_{1}$ as
\begin{align}
u_{0}=w_{0}-\frac{w'_{1}}{2}-\frac{(w_{1})^{2}}{4}=-\frac{w''_{1}}{
 2w_{1}}+\frac{(w'_{1})^{2}}{4(w_{1})^{2}}+\frac{4C_{1}}{(w_{1})^{2}}.
\label{eq:2fu0}
\end{align}
Substituting it into (\ref{eq:P2+-}) and (\ref{eq:V2+-}), we finally get
the general form of $2$-fold SUSY systems as
\begin{align}
P_{2}^{\pm}&\stackrel{2}{\sim}\del^{2}\mp w_{1}\del+\frac{(w_{1})^{2}}{4}
 -\frac{w''_{1}}{2w_{1}}+\frac{(w'_{1})^{2}}{4(w_{1})^{2}}+\frac{4C_{1}}{(w_{1})^{2}}
 \mp\frac{w'_{1}}{2},\\
V^{\pm}&\stackrel{2}{\sim}\frac{(w_{1})^{2}}{8}+\frac{w''_{1}}{4w_{1}}
 -\frac{(w'_{1})^{2}}{8(w_{1})^{2}}-\frac{2C_{1}}{(w_{1})^{2}}\pm\frac{w'_{1}}{2}-C_{0}.
\end{align}
Finally, products of the components of $2$-fold supercharges $P_{2}^{\mp}
P_{2}^{\pm}$ are calculated as
\begin{align}
P_{2}^{-}P_{2}^{+}&=4[(H^{+}+C_{0})^{2}+C_{1}]-2I_{0},\\
P_{2}^{+}P_{2}^{-}&=4[(H^{-}+C_{0})^{2}+C_{1}]+2I_{0},
\end{align}
where $I_{0}$ and $C_{1}$ are given by (\ref{eq:2fc3'}) and (\ref{eq:2fC1}),
respectively. Hence, we obtain the equality (\ref{eq:PNPN}) for $\cN=2$
as an equivalent relation associated with $2$-fold SUSY:
\begin{align}
P_{2}^{\mp}P_{2}^{\pm}\stackrel{2}{\sim}4[(H^{\pm}+C_{0})^{2}+C_{1}].
\end{align}

\section{3-fold Supersymmetry}
\label{sec:3f}

Next, we shall reexamine the case of $3$-fold SUSY, which was once investigated
briefly in Ref.~\cite{IN04}, by utilizing our general analysis. Components of $3$-fold 
supercharges are given by
\begin{align}
\begin{split}
P_{3}^{-}&=\del^{3}+w_{2}\del^{2}+w_{1}\del+w_{0},\\
P_{3}^{+}&=-\del^{3}+w_{2}\del^{2}-(w_{1}-2w'_{2})\del+w_{0}-w'_{1}+w''_{2}.
\label{eq:3fP}
\end{split}
\end{align}
The condition for $3$-fold SUSY $P_{3}^{-}H^{-}-H^{+}P_{3}^{-}=0$ is
satisfied if and only if the following four equalities hold:
\begin{align}
&V^{+}-V^{-}=w'_{2},
\label{eq:3fc1}\\
&w''_{2}+2w'_{1}+6V^{-\prime}-2w_{2}(V^{+}-V^{-})=0,
\label{eq:3fc2}\\
&w''_{1}+2w'_{0}+6V^{-\pprime}+4w_{2}V^{-\prime}-2w_{1}(V^{+}-V^{-})=0,
\label{eq:3fc3}\\
&w''_{0}+2V^{-\ppprime}+2w_{2}V^{-\pprime}+2w_{1}
 V^{-\prime}-2w_{0}(V^{+}-V^{-})=0.
\label{eq:3fc4}
\end{align}
Substituting (\ref{eq:3fc1}) into (\ref{eq:3fc2})--(\ref{eq:3fc4}) and integrating
the resulting equation from (\ref{eq:3fc2}), we obtain
\begin{align}
6V^{+}=&\;5w'_{2}-2w_{1}+(w_{2})^{2}-6C_{0},\\
6V^{-}=&\;-w'_{2}-2w_{1}+(w_{2})^{2}-6C_{0},\\
-6I_{1}=&\;3w'''_{2}+3w''_{1}-6w'_{0}-4w_{2}w''_{2}-6(w'_{2})^{2}+6w'_{2}w_{1}
 +4w_{2}w'_{1}-4(w_{2})^{2}w'_{2}=0,
\label{eq:3fc3'}\\
-6I_{0}=&\;w''''_{2}+2w'''_{1}-3w''_{0}-w_{2}w'''_{2}-6w'_{2}w''_{2}+w''_{2}w_{1}
 +2w_{2}w''_{1}+6w'_{2}w_{0}+2w_{1}w'_{1}\notag\\
&\;-2(w_{2})^{2}w''_{2}-2w_{2}(w'_{2})^{2}-2w_{2}w'_{2}w_{1}=0,
\label{eq:3fc4'}
\end{align}
where $C_{0}$ is an integral constant. To integrate the remaining equations
(\ref{eq:3fc3'}) and (\ref{eq:3fc4'}), we shall first construct a set of
dimension-preserving transformations $w_{k}\to u_{k}$ ($k=0,1$) which will
convert simultaneously both the pairs $P_{3}^{\pm}$ and $V^{\pm}$ into
symmetric forms, like (\ref{eq:2ftf}) in the case of $2$-fold SUSY. The most
general transformations of polynomial type preserving the physical dimensions
$[\rmL^{k-3}]$ of $w_{k}$ would be
\begin{align}
\begin{split}
w_{1}&=6u_{1}+w'_{2}-\alpha_{1}(w_{2})^{2}\ [\rmL^{-2}],\\
w_{0}&=u_{0}+3u'_{1}-\beta_{1}w''_{2}-\alpha_{1}w_{2}w'_{2}-6\beta_{2}
 w_{2}u_{1}-\beta_{3}(w_{2})^{3}\ [\rmL^{-3}],
\label{eq:3ftf}
\end{split}
\end{align}
where $u_{k}$ $[\rmL^{k-3}]$ and $\alpha_{0}$, $\beta_{k}$ ($k=1,2,3$) are
all dimensionless parameters. They actually render both the pair of $3$-fold
supercharge components and the pair of potentials of symmetric forms as
\begin{subequations}
\label{eqs:3fPV}
\begin{align}
P_{3}^{\pm}=&\;\mp\del^{3}+w_{2}\del^{2}\mp[6u_{1}-\alpha_{1}(w_{2})^{2}
 \mp w'_{2}]\del\notag\\
&\;+u_{0}-\beta_{1}w''_{2}-6\beta_{2}w_{2}u_{1}-\beta_{3}(w_{2})^{3}
 \mp(3u'_{1}-\alpha_{1}w_{2}w'_{2}),\\
6V^{\pm}=&\;-12u_{1}+(2\alpha_{1}+1)(w_{2})^{2}\pm 3w'_{2}-6C_{0}.
\end{align}
\end{subequations}
With the transformations (\ref{eq:3ftf}), the two constraints (\ref{eq:3fc3'})
and (\ref{eq:3fc4'}) are equivalent to the following new set of conditions:
\begin{align}
-3\bar{I}_{1}=&\;9I_{1}\notag\\
=&\;3(\beta_{1}+1)w'''_{2}-3u'_{0}+18(\beta_{2}+1)w'_{2}u_{1}
 +6(3\beta_{2}+2)w_{2}u'_{1}\notag\\
&\;-(7\alpha_{1}-9\beta_{3}+2)(w_{2})^{2}w'_{2}=0\ [\rmL^{-4}],
\label{eq:3fc3g}\\
3\bar{I}_{0}=&\;-9(2I_{0}-\del I_{1})\notag\\
=&\;3u'''_{1}-(\alpha_{1}-1)w_{2}w'''_{2}-3(2\beta_{1}+\alpha_{1}+1)
 w'_{2}w''_{2}+6w'_{2}u_{0}\notag\\
&\;+72u_{1}u'_{1}-12(2\alpha_{1}+3\beta_{2}+1)w_{2}w'_{2}u_{1}-12\alpha_{1}
 (w_{2})^{2}u'_{1}\notag\\
&\;+2[2(\alpha_{1})^{2}+\alpha_{1}-3\beta_{3}](w_{2})^{3}w'_{2}=0\ [\rmL^{-5}].
\label{eq:3fc4g}
\end{align}
Hence, they would get simplest when
\begin{align}
\alpha_{1}=1,\quad\beta_{1}=-1,\quad\beta_{2}=-1,\quad\beta_{3}=1.
\end{align}
With the latter choice of parameter values, the new set of conditions
(\ref{eq:3fc3g}) and (\ref{eq:3fc4g}) reads as
\begin{align}
&\bar{I}_{1}=u'_{0}+2w_{2}u'_{1}=0\ [\rmL^{-4}],
\label{eq:3fc3''}\\
&\bar{I}_{0}=u'''_{1}+2w'_{2}u_{0}+24u_{1}u'_{1}-4(w_{2})^{2}u'_{1}=0\ [\rmL^{-5}].
\label{eq:3fc4''}
\end{align}
The first equation (\ref{eq:3fc3''}) enables us to express $u_{0}$ in terms of
$u_{1}$ and $w_{2}$ as an indefinite integral
\begin{align}
u_{0}=-2\int\rmd q\,w_{2}u'_{1}\ [\rmL^{-3}].
\end{align}
Next, let us construct the total differential $\rmd J_{1}/\rmd q$ in (\ref{eq:dJdq2}).
We note that $\rmd J_{1}/\rmd q$ and $\bar{I}_{k}$ ($k=1,2$) have the physical
dimensions $[\rmL^{-5}]$ and $[\rmL^{k-5}]$, respectively. Hence, the operators
$L_{1k}$ defined in (\ref{eq:dJdq2}) in this case must have the physical dimension 
$[\rmL^{-k}]$. Thus, $L_{10}$ is just a dimensionless constant while $L_{11}\propto
w_{2}$ if we restrict $L_{1k}$ to multiplicative operators of polynomial type. Indeed,
we can easily find that one of the latter choices leads to a total differential
\begin{align}
-4\frac{\rmd J_{1}}{\rmd q}=\bar{I}_{0}+2w_{2}\bar{I}_{1}=
 u'''_{1}+2(w'_{2}u_{0}+w_{2}u'_{0})+24u_{1}u'_{1}=0\ [\rmL^{-5}],
\end{align}
which can be easily integrated as
\begin{align}
-4J_{1}[w]=u''_{1}+2w_{2}u_{0}+12(u_{1})^{2}=-4C_{1}\ [\rmL^{-4}],
\label{eq:3fC1}
\end{align}
where $C_{1}$ is another integral constant having the correct physical
dimension $[\rmL^{-4}]$ listed in (\ref{eq:pdim1}). From (\ref{eq:3fc4''}) and
(\ref{eq:3fC1}), we can express $u_{0}$ in terms of $u_{1}$ and $w_{2}$
without recourse to any indefinite integral as
\begin{align}
u_{0}=-\frac{u'''_{1}+24u_{1}u'_{1}-4(w_{2})^{2}u'_{1}}{2w'_{2}}
 =-\frac{u''_{1}+12(u_{1})^{2}+4C_{1}}{2w_{2}}.
\label{eq:3fu0}
\end{align}
To construct the second integral $C_{2}$ of $3$-fold SUSY systems, we first
note that $\rmd J_{2}/\rmd q$ and $\bar{I}_{k}$ ($k=0,1$) have the physical
dimensions $[\rmL^{-7}]$ and $[\rmL^{k-5}]$, respectively. Hence, the operators
$L_{2k}$ defined in (\ref{eq:dJdq2}) in this case must have the physical dimension 
$[\rmL^{-k-2}]$. Thus, candidates for $L_{20}$ are $w'_{2}$, $u_{1}$, and $(w_{2})^{2}$,
while those for $L_{21}$ are $w''_{2}$, $u'_{1}$, $u_{0}$, $w_{2}w'_{2}$, $w_{2}u_{1}$, and 
$(w_{2})^{3}$, if we restrict $L_{2k}$ to multiplicative operators of polynomial type.
With the choice of $L_{20}=0$ and $L_{21}\propto u_{0}$, we can construct
a total differential as
\begin{align}
4\frac{\rmd J_{2}}{\rmd q}=u_{0}\bar{I}_{1}&=u_{0}u'_{0}+2w_{2}u'_{1}u_{0}\notag\\
&=u_{0}u'_{0}-u'_{1}u''_{1}-12(u_{1})^{2}u'_{1}-4C_{1}u'_{1}=0\ [\rmL^{-7}],
\end{align}
where (\ref{eq:3fC1}) has been used.
The latter relation is indeed easily integrated as
\begin{align}
8J_{2}[w]=(u_{0})^{2}-(u'_{1})^{2}-8(u_{1})^{3}-8C_{1}u_{1}=8C_{2}\ [\rmL^{-6}],
\label{eq:3fC2}
\end{align}
with another integral constant $C_{2}$ having the correct physical
dimension $[\rmL^{-6}]$ listed in (\ref{eq:pdim1}). With the use of (\ref{eq:3fC2})
we can express $u_{0}$ solely in terms of $u_{1}$. Then, substituting the obtained
expression for $u_{0}$ into (\ref{eq:3fc3''}), we can also express $w_{2}$ solely in
terms of $u_{1}$. Hence, we can eventually have an expression for $V^{\pm}$
and $P_{3}^{\pm}$ in terms of only a single arbitrary function $u_{1}$. However,
the latter expression is relatively complicated and thus it would be more
convenient in practice to express them in terms of two of the three functions
$w_{2}$, $u_{1}$, and $u_{0}$. If we eliminate $u_{0}$ in (\ref{eqs:3fPV}) by using
(\ref{eq:3fu0}), we have the expression in terms of $w_{2}$ and $u_{1}$ as
\begin{subequations}
\begin{align}
P_{3}^{\pm}\stackrel{3}{\sim}&\;\mp\del^{3}+w_{2}\del^{2}\mp\left[6u_{1}
 -(w_{2})^{2}\mp w'_{2}\right]\del\notag\\
&\;+w''_{2}+6w_{2}u_{1}-(w_{2})^{3}-\frac{u''_{1}}{2w_{2}}-\frac{6(u_{1})^{2}}{w_{2}}
 -\frac{2C_{1}}{w_{2}}\mp (3u'_{1}-w_{2}w'_{2}),\\
V^{\pm}=&\;-2u_{1}+\frac{1}{2}(w_{2})^{2}\pm\frac{1}{2}w'_{2}-C_{0}.
\end{align}
\end{subequations}
On the other hand, if we eliminate $w_{2}$ in (\ref{eqs:3fPV}) by using
(\ref{eq:3fc3''}), we have the expression in terms of $u_{1}$ and $u_{0}$ as
\begin{subequations}
\begin{align}
P_{3}^{\pm}\stackrel{3}{\sim}&\;\mp\del^{3}-\frac{u'_{0}}{2u'_{1}}\del^{2}\mp
 \left[6u_{1}-\frac{(u'_{0})^{2}}{4(u'_{1})^{2}}\pm\left(\frac{u''_{0}}{2u'_{1}}
 -\frac{u''_{1}u'_{0}}{2(u'_{1})^{2}}\right)\right]\del+u_{0}
 -\frac{u'''_{0}}{2u'_{1}}+\frac{u''_{1}u''_{0}}{(u'_{1})^{2}}\notag\\
&\;+\left(\frac{u'''_{1}}{2(u'_{1})^{2}}-\frac{(u''_{1})^{2}}{(u'_{1})^{3}}
 -\frac{3u_{1}}{u'_{1}}\right)u'_{0}+\frac{(u'_{0})^{3}}{8(u_{1})^{3}}
 \mp\left(3u'_{1}-\frac{u'_{0}u''_{0}}{4(u'_{1})^{2}}+\frac{u''_{1}
 (u'_{0})^{2}}{4(u'_{1})^{3}}\right),\\
V^{\pm}\stackrel{3}{\sim}&\;-2u_{1}+\frac{(u'_{0})^{2}}{8(u'_{1})^{2}}\mp\left(
 \frac{u''_{0}}{4u'_{1}}-\frac{u''_{1}u'_{0}}{4(u'_{1})^{2}}\right)-C_{0}.
\end{align}
\end{subequations}
Finally, products of the components of $3$-fold supercharges $P_{3}^{\mp}
P_{3}^{\pm}$ are calculated as
\begin{align}
P_{3}^{-}P_{3}^{+}=&\;8[(H^{+}+C_{0})^{3}+C_{1}(H^{+}+C_{0})+C_{2}]+3I_{1}\del^{2}
 +2[(2\del+w_{2})I_{1}-I_{0}]\del\notag\\
&\;+(2\del^{2}+2w_{2}\del-2w'_{2}+w_{1})I_{1}-2(\del+w_{2})I_{0},\\
P_{3}^{+}P_{3}^{-}=&\;8[(H^{-}+C_{0})^{3}+C_{1}(H^{-}+C_{0})+C_{2}]-3I_{1}\del^{2}
 -2[(\del-w_{2})I_{1}+I_{0}]\del\notag\\
&\;-w_{1}I_{1}-2(\del-w_{2})I_{0},
\end{align}
where $I_{1}$, $I_{0}$, $C_{1}$, and $C_{2}$ are given by (\ref{eq:3fc3'}),
(\ref{eq:3fc4'}), (\ref{eq:3fC1}), and (\ref{eq:3fC2}),
respectively. Hence, we obtain the equality (\ref{eq:PNPN}) for $\cN=3$
as an equivalent relation associated with $3$-fold SUSY:
\begin{align}
P_{3}^{\mp}P_{3}^{\pm}\stackrel{3}{\sim}8[(H^{\pm}+C_{0})^{3}
 +C_{1}(H^{\pm}+C_{0})+C_{2}].
\end{align}

\section{4-fold Supersymmetry}
\label{sec:4f}

In this section, we shall study the case of $4$-fold supersymmetry.
Components of $4$-fold supercharges are given by
\begin{align}
\begin{split}
P_{4}^{-}&=\del^{4}+w_{3}\del^{3}+w_{2}\del^{2}+w_{1}\del+w_{0},\\
P_{4}^{+}&=\del^{4}-w_{3}\del^{3}+(w_{2}-3w'_{3})\del^{2}
 -(w_{1}-2w'_{2}+3w''_{3})\del+w_{0}-w'_{1}+w''_{2}-w'''_{3}.
\label{eq:4fP}
\end{split}
\end{align}
The condition for $4$-fold supersymmetry $P_{4}^{-}H^{-}-H^{+}P_{4}^{-}=0$ is
satisfied if and only if the following five equalities hold:
\begin{align}
&V^{+}-V^{-}=w'_{3},
\label{eq:4fc1}\\
&w''_{3}+2w'_{2}+8V^{-\prime}-2w_{3}(V^{+}-V^{-})=0,
\label{eq:4fc2}\\
&w''_{2}+2w'_{1}+12V^{-\pprime}+6w_{3}V^{-\prime}-2w_{2}(V^{+}-V^{-})=0,
\label{eq:4fc3}\\
&w''_{1}+2w'_{0}+8V^{-\ppprime}+6w_{3}V^{-\pprime}
 +4w_{2}V^{-\prime}-2w_{1}(V^{+}-V^{-})=0,
\label{eq:4fc4}\\
&w''_{0}+2V^{-\ppprime\prime}+2w_{3}V^{-\ppprime}
 +2w_{2}V^{-\pprime}+2w_{1}V^{-\prime}-2w_{0}(V^{+}-V^{-})=0.
\label{eq:4fc5}
\end{align}
Substituting (\ref{eq:4fc1}) into (\ref{eq:4fc2})--(\ref{eq:4fc5}) and integrating
the resulting equation from (\ref{eq:4fc2}), we obtain
\begin{align}
8V^{+}=&\;7w'_{3}-2w_{2}+(w_{3})^{2}-8C_{0},\\
8V^{-}=&\;-w'_{3}-2w_{2}+(w_{3})^{2}-8C_{0},\\
-8I_{2}=&\;6w'''_{3}+8w''_{2}-8w'_{1}-9w_{3}w''_{3}-12(w'_{3})^{2}+8w'_{3}w_{2}
 +6w_{3}w'_{2}-6(w_{3})^{2}w'_{3}=0,
\label{eq:4fc3'}\\
-8I_{1}=&\;4w''''_{3}+8w'''_{2}-4w''_{1}-8w'_{0}-5w_{3}w'''_{3}-24w'_{3}w''_{3}
 +2w''_{3}w_{2}+6w_{3}w''_{2}\notag\\
&\;+8w'_{3}w_{1}+4w_{2}w'_{2}-6(w_{3})^{2}w''_{3}-6w_{3}(w'_{3})^{2}
 -4w_{3}w'_{3}w_{2}=0,
\label{eq:4fc4'}\\
-8I_{0}=&\;w'''''_{3}+2w''''_{2}-4w''_{0}-w_{3}w''''_{3}-8w'_{3}w'''_{3}
 -6(w''_{3})^{2}+w'''_{3}w_{2}+2w_{3}w'''_{2}\notag\\
&\;+w''_{3}w_{1}+8w'_{3}w_{0}+2w_{2}w''_{2}+2w'_{2}w_{1}-2(w_{3})^{2}w'''_{3}
 -6w_{3}w'_{3}w''_{3}\notag\\
&\;-2w_{3}w''_{3}w_{2}-2(w'_{3})^{2}w_{2}-2w_{3}w'_{3}w_{1}=0.
\label{eq:4fc5'}
\end{align}
where $C_{0}$ is an integral constant. To integrate the remaining equations
(\ref{eq:4fc3'})--(\ref{eq:4fc5'}), let us first look for a set of dimension-preserving
transformations $w_{k}\to u_{k}$ ($k=0,1,2$) which will convert simultaneously
both the pairs $P_{4}^{\pm}$ and $V^{\pm}$ into symmetric forms as in the cases
of $2$- and $3$-fold SUSY. The most general transformations of polynomial type
preserving the physical dimensions $[\rmL^{k-4}]$ of $w_{k}$ would be
\begin{align}
w_{2}=&\;u_{2}+\frac{3}{2}w'_{3}-\alpha_{1}(w_{3})^{2}\ [\rmL^{-2}],\\
w_{1}=&\;u_{1}+u'_{2}-\beta_{1}w''_{3}-2\alpha_{1}w_{3}w'_{3}-\beta_{2}w_{3}u_{2}
 -\beta_{3}(w_{3})^{3}\ [\rmL^{-3}],\\
w_{0}=&\;u_{0}+\frac{1}{2}u'_{1}-\gamma_{1}u''_{2}-\left(\frac{\beta_{1}}{2}
 +\frac{1}{4}\right)w'''_{3}-\gamma_{2}w_{3}w''_{3}-\gamma_{3}(w'_{3})^{2}
 -\frac{\beta_{2}}{2}(w_{3}u_{2})'\notag\\
&\;-\gamma_{4}w_{3}u_{1}-\gamma_{5}(u_{2})^{2}-\frac{3\beta_{3}}{2}(w_{3})^{2}w'_{3}
 -\gamma_{6}(w_{3})^{2}u_{2}+\gamma_{7}(w_{3})^{4}\ [\rmL^{-4}],
\label{eq:4ftf}
\end{align}
where $u_{k}$ $[\rmL^{k-4}]$ and $\alpha_{0}$, $\beta_{k}$ ($k=1,2,3$), and
$\gamma_{k}$ ($k=1,\dots,7$) are all dimensionless parameters. They indeed
make both the pair of $4$-fold supercharge components and the pair of
potentials symmetric as
\begin{subequations}
\label{eqs:4fPV}
\begin{align}
P_{4}^{\pm}=&\;\del^{4}\mp w_{3}\del^{3}+\left[u_{2}-\alpha_{1}(w_{3})^{2}
 \mp\frac{3}{2}w'_{3}\right]\del^{2}\mp\biggl[u_{1}-\beta_{1}w''_{3}
 -\beta_{2}w_{3}u_{2}-\beta_{3}(w_{3})^{3}\notag\\
&\;\mp(u'_{2}-2\alpha_{1}w_{3}w'_{3})\biggr]\del+u_{0}-\gamma_{1}u''_{2}
 -\gamma_{2}w_{3}w''_{3}-\gamma_{3}(w'_{3})^{2}-\gamma_{4}w_{3}u_{1}
 -\gamma_{5}(u_{2})^{2}\notag\\
&\;-\gamma_{6}(w_{3})^{2}u_{2}-\gamma_{7}(w_{3})^{4}
 \mp\left[\frac{1}{2}u'_{1}-\left(\frac{\beta_{1}}{2}+\frac{1}{4}\right)w'''_{3}
 -\frac{\beta_{2}}{2}(w_{3}u_{2})'-\frac{3\beta_{3}}{2}(w_{3})^{2}w'_{3}\right],\\
8V^{\pm}=&\;-2u_{2}+(2\alpha_{1}+1)(w_{3})^{2}\pm 4w'_{3}-8C_{0}.
\end{align}
\end{subequations}
With the transformations (\ref{eq:4ftf}), the remaining three constraints
(\ref{eq:4fc3'})--(\ref{eq:4fc5'}) are equivalent to the following new set of
conditions:
\begin{align}
\bar{I}_{2}=&\;4I_{2}\notag\\
=&\;-(4\beta_{1}+9)w'''_{3}+4u'_{1}-4(\beta_{2}+1)w'_{3}u_{2}-(4\beta_{2}
 +3)w_{3}u'_{2}\notag\\
&\;+(10\alpha_{1}-12\beta_{3}+3)(w_{3})^{2}w'_{3}=0\ [\rmL^{-4}],
\label{eq:4fc3g}
\end{align}
\begin{align}
4\bar{I}_{1}=&\;4(I_{1}-\del I_{2})\notag\\
=&\;-2(2\gamma_{1}+1)u'''_{2}+4u'_{0}-(4\beta_{1}\gamma_{4}-4\alpha_{1}
 +4\gamma_{2}+9\gamma_{4}+2)w_{3}w'''_{3}\notag\\
&\;+2(6\alpha_{1}+2\beta_{1}-2\gamma_{2}-4\gamma_{3}+3)w'_{3}w''_{3}
 -4(\gamma_{4}+1)w'_{3}u_{1}\notag\\
&\;-2(4\gamma_{5}+1)u_{2}u'_{2}-2(2\beta_{2}\gamma_{4}-2\alpha_{1}-2\beta_{2}
 +2\gamma_{4}+4\gamma_{6}-1)w_{3}w'_{3}u_{2}\notag\\
&\;-(4\beta_{2}\gamma_{4}-2\alpha_{1}+3\gamma_{4}+4\gamma_{6})
 (w_{3})^{2}u'_{2}\notag\\
&\;-[4(\alpha_{1})^{2}-10\alpha_{1}\gamma_{4}+12\beta_{3}\gamma_{4}+2\alpha_{1}
 -3\gamma_{4}-4\beta_{3}+16\gamma_{7}](w_{3})^{3}w'_{3}=0\ [\rmL^{-5}],
\label{eq:4fc4g}
\end{align}
\begin{align}
\bar{I}_{0}=&\;-4(4I_{0}-2\del I_{1}+\del^{2}I_{2})\notag\\
=&\;w'''''_{3}+4w'''_{3}u_{2}-(4\beta_{1}+3)w''_{3}u'_{2}-4(4\gamma_{1}+1)w'_{3}u''_{2}
 +w_{3}u'''_{2}+16w'_{3}u_{0}\notag\\
&\;+4u'_{2}u_{1}-(6\alpha_{1}+1)(w_{3})^{2}w'''_{3}+4(2\alpha_{1}\beta_{1}
 +2\alpha_{1}+\beta_{1}-4\gamma_{2})w_{3}w'_{3}w''_{3}\notag\\
&\;+8(\alpha_{1}-2\gamma_{3})(w'_{3})^{3}-4(2\alpha_{1}+4\gamma_{4}+1)w_{3}w'_{3}
 u_{1}-16\gamma_{5}w'_{3}(u_{2})^{2}\notag\\
&\;-4\beta_{2}w_{3}u_{2}u'_{2}+4(2\alpha_{1}\beta_{2}+\beta_{2}-4\gamma_{6})
 (w_{3})^{2}w'_{3}u_{2}-4\beta_{3}(w_{3})^{3}u'_{2}\notag\\
&\;+4(2\alpha_{1}\beta_{3}+\beta_{3}-4\gamma_{7})(w_{3})^{4}w'_{3}=0\ [\rmL^{-6}].
\label{eq:4fc5g}
\end{align}
Hence, they would get simplest when\footnote{Another possible choice could be
$\alpha_{1}=0$, $\beta_{1}=-9/4$, $\beta_{2}=-3/4$, $\beta_{3}=1/4$,
$\gamma_{1}=-1/2$, $\gamma_{2}=-1/2$, $\gamma_{3}=-1/8$, $\gamma_{4}=-1$,
$\gamma_{5}=-1/4$, $\gamma_{6}=0$, $\gamma_{7}=1/16$. With the latter
choice, it turns out that the third integral $C_{3}$ admits a simpler expression
than (\ref{eq:4fC3}) but the expression for the second integral $C_{2}$ contains
terms which are linear in $u_{1}$, in contrast with (\ref{eq:4fC2}) and
(\ref{eq:4fu1'}).}
\begin{align}
\begin{split}
&\alpha_{1}=3/2,\quad\beta_{1}=-9/4,\quad\beta_{2}=-1,\quad\beta_{3}=3/2,
 \quad\gamma_{1}=-1/2,\quad\gamma_{2}=1,\\
&\gamma_{3}=11/8,\quad\gamma_{4}=-1,\quad\gamma_{5}=-1/4,\quad
 \gamma_{6}=1/2,\quad\gamma_{7}=-3/8.
\label{eq:4fpv1}
\end{split}
\end{align}
With the latter choice of parameter values, the new set of conditions
(\ref{eq:4fc3g})--(\ref{eq:4fc5g}) reads as
\begin{align}
\bar{I}_{2}=&\;4u'_{1}+w_{3}u'_{2}=0\ [\rmL^{-4}],
\label{eq:4fc3''}\\
\bar{I}_{1}=&\;u'_{0}=0\ [\rmL^{-5}],
\label{eq:4fc4''}\\
\bar{I}_{0}=&\;w'''''_{3}+4w'''_{3}u_{2}+6w''_{3}u'_{2}+4w'_{3}u''_{2}+w_{3}u'''_{2}
 +16w'_{3}u_{0}+4u'_{2}u_{1}\notag\\
&\;-10(w_{3})^{2}w'''_{3}-40w_{3}w'_{3}w''_{3}-10(w'_{3})^{3}+4w'_{3}(u_{2})^{2}
 +4w_{3}u_{2}u'_{2}\notag\\
&\;-24(w_{3})^{2}w'_{3}u_{2}-6(w_{3})^{3}u'_{2}+30(w_{3})^{4}w'_{3}=0\ [\rmL^{-6}].
\label{eq:4fc5''}
\end{align}
The first equation (\ref{eq:4fc3''}) enables us to express $u_{1}$ in terms of
$u_{2}$ and $w_{3}$ as an indefinite integral
\begin{align}
4u_{1}=-\int\rmd q\,w_{3}u'_{2}\ [\rmL^{-3}].
\end{align}
The second equation (\ref{eq:4fc4''}) just means that $u_{0}$ is a constant.
On the other hand, $C_{k}$ ($k=0,\dots,\cN-1$) are the only constants which
appear in general form of $\cN$-fold SUSY. The physical dimension of $u_{0}$
is $[\rmL^{-4}]$ and thus we can put
\begin{align}
u_{0}=2C_{1}\ [\rmL^{-4}],
\label{eq:4fC1}
\end{align}
where $C_{1}$ is an integral constant having the correct physical dimension
$[\rmL^{-4}]$ listed in (\ref{eq:pdim1}). We have now obtained the first integral
$C_{1}$ and thus can skip over the step to construct the quantity $J_{1}$.

Next, to construct the second integral $C_{2}$ of $4$-fold SUSY systems, we first
note that $\rmd J_{2}/\rmd q$ and $\bar{I}_{k}$ ($k=0,1,2$) have the physical
dimensions $[\rmL^{-7}]$ and $[\rmL^{k-6}]$, respectively. Hence, the operators
$L_{2k}$ defined in (\ref{eq:dJdq2}) in this case must have the physical dimension 
$[\rmL^{-k-1}]$. Thus, candidates for $L_{20}$ are $w_{3}$, those for $L_{21}$ are
$w'_{3}$, $u_{2}$, and $(w_{3})^{2}$, while those for $L_{22}$ are $w''_{3}$, $u'_{2}$,
$u_{1}$, $w_{3}w'_{3}$, $w_{3}u_{2}$, and $(w_{3})^{3}$, if we restrict $L_{2k}$ to
multiplicative operators of polynomial type. With the choice of $L_{20}\propto w_{3}$
and $L_{21}=L_{22}=0$, we can construct a total differential as
\begin{align}
-128\frac{\rmd J_{2}}{\rmd q}=2w_{3}\bar{I}_{0}=0\ [\rmL^{-7}],
\end{align}
where we have omitted the explicit expression. The integration of the latter yields
\begin{align}
-128J_{2}[w]=&\;2w_{3}w''''_{3}-2w'_{3}w'''_{3}+(w''_{3})^{2}-16(u_{1})^{2}
 +8w_{3}w''_{3}u_{2}-4(w'_{3})^{2}u_{2}\notag\\
&\;+4w_{3}w'_{3}u'_{2}+2(w_{3})^{2}u''_{2}-20(w_{3})^{3}w''_{3}-10(w_{3})^{2}(w'_{3})^{2}
 +4(w_{3})^{2}(u_{2})^{2}\notag\\
&\;-12(w_{3})^{4}u_{2}+10(w_{3})^{6}+32C_{1}(w_{3})^{2}=-128C_{2}\ [\rmL^{-6}],
\label{eq:4fC2}
\end{align}
where (\ref{eq:4fC1}) has been applied, and
$C_{2}$ is another integral constant having the correct physical
dimension $[\rmL^{-6}]$ listed in (\ref{eq:pdim1}). From (\ref{eq:4fc5''}) and
(\ref{eq:4fC1}), we can express $u_{1}$ in terms of $u_{2}$ and $w_{3}$
without recourse to any indefinite integral as
\begin{align}
u_{1}=&\;[-w'''''_{3}-4w'''_{3}u_{2}-6w''_{3}u'_{2}-4w'_{3}u''_{2}
 -w_{3}u'''_{2}+10(w_{3})^{2}w'''_{3}\notag\\
&\;+40w_{3}w'_{3}w''_{3}+10(w'_{3})^{3}-4w'_{3}(u_{2})^{2}-4w_{3}u_{2}u'_{2}
 +24(w_{3})^{2}w'_{3}u_{2}\notag\\
&\;+6(w_{3})^{3}u'_{2}-30(w_{3})^{4}w'_{3}-32C_{1}w'_{3}]/(4u'_{2}).
\label{eq:4fu1}
\end{align}
Instead, using (\ref{eq:4fC2}), we can express $u_{1}$ in terms of $u_{2}$ and
$w_{3}$ as a solution to the following quadratic equation
\begin{align}
16(u_{1})^{2}=&\;2w_{3}w''''_{3}-2w'_{3}w'''_{3}+(w''_{3})^{2}+8w_{3}w''_{3}u_{2}
 -4(w'_{3})^{2}u_{2}+4w_{3}w'_{3}u'_{2}\notag\\
&\;+2(w_{3})^{2}u''_{2}-20(w_{3})^{3}w''_{3}-10(w_{3})^{2}(w'_{3})^{2}
 +4(w_{3})^{2}(u_{2})^{2}\notag\\
&\;-12(w_{3})^{4}u_{2}+10(w_{3})^{6}+32C_{1}(w_{3})^{2}+128C_{2}.
\label{eq:4fu1'}
\end{align}
To obtain the third integral $C_{3}$ of $4$-fold SUSY systems, we
note that $\rmd J_{3}/\rmd q$ and $\bar{I}_{k}$ ($k=0,1,2$) have the physical
dimensions $[\rmL^{-9}]$ and $[\rmL^{k-6}]$, respectively. Hence, the operators
$L_{3k}$ defined in (\ref{eq:dJdq2}) in this case must have the physical dimension 
$[\rmL^{-k-3}]$. With a similar dimensonal analysis, we find that the choice of 
$L_{30}\propto w''_{3}+4u_{1}+2w_{3}u_{2}-2(w_{3})^{3}$ and $L_{31}=L_{32}=0$
leads to a total differential:
\begin{align}
512\frac{\rmd J_{3}}{\rmd q}=
[w''_{3}+4u_{1}+2w_{3}u_{2}-2(w_{3})^{3}]\bar{I}_{0}=0\ [\rmL^{-9}],
\end{align}
where we have omitted the explicit expression.
The integral of the latter equation reads as $J_{3}[w]=C_{3}$
with another integral constant $C_{3}$ having the correct physical dimension
$[\rmL^{-8}]$ listed in (\ref{eq:pdim1}). The explicit formula for $J_{3}$ is
given by (\ref{eq:4fC3}) in Appendix~A. By using (at least) two of the equalities
(\ref{eq:4fu1}), (\ref{eq:4fu1'}), and (\ref{eq:4fC3}), we can eliminate $u_{1}$
to obtain the relation between $w_{3}$ and $u_{2}$. Hence, we are now, in
principle, able to express a $4$-fold SUSY system solely in terms of a single
arbitrary function. As in the case of $3$-fold SUSY, however, it would be
more convenient for a practical purpose to have an expression in terms of
two of the four functions $w_{3}$, $u_{2}$, $u_{1}$, and $u_{0}$. For example,
if we eliminate $u_{1}$ and $u_{0}$ in the system by using (\ref{eq:4fu1}) and
(\ref{eq:4fC1}), respectively, the system (\ref{eqs:4fPV}) with the parameter
values (\ref{eq:4fpv1}) can be represented in terms of $w_{3}$ and $u_{2}$ as
(\ref{eqs:4fPV'}).

Finally, products of the components of $4$-fold supercharges $P_{4}^{\mp}
P_{4}^{\pm}$ are calculated as
\begin{align}
P_{4}^{-}P_{4}^{+}=&\;16[(H^{+}+C_{0})^{4}+C_{1}(H^{+}+C_{0})^{2}+C_{2}(H^{+}+C_{0})
 +C_{3}]+\sum_{i=0}^{4}f_{i}^{+}[w]\del^{i},
\label{eq:P4-+}\\
P_{4}^{+}P_{4}^{-}=&\;16[(H^{-}+C_{0})^{4}+C_{1}(H^{-}+C_{0})^{2}+C_{2}(H^{-}+C_{0})
 +C_{3}]+\sum_{i=0}^{4}f_{i}^{-}[w]\del^{i},
\label{eq:P4+-}
\end{align}
where $C_{1}$, $C_{2}$, and $C_{3}$ are given by (\ref{eq:4fC1}), (\ref{eq:4fC2}), and
(\ref{eq:4fC3}), respectively. Both of the second terms in the above are elements
of $\cK_{0}^{[4]}$ introduced in Section~\ref{ssec:eqc} and have the form of
(\ref{eq:K0}). The explicit forms of $f_{i}^{\pm}$ are given by (\ref{eq:f+}) and
(\ref{eq:f-}). Hence, we obtain the equality (\ref{eq:PNPN})
for $\cN=4$ as an equivalent relation associated with $4$-fold SUSY:
\begin{align}
P_{4}^{\mp}P_{4}^{\pm}\stackrel{4}{\sim}16[(H^{\pm}+C_{0})^{4}
 +C_{1}(H^{\pm}+C_{0})^{2}+C_{2}(H^{\pm}+C_{0})+C_{3}].
\end{align}

\section{Discussion and Summary}
\label{sec:discus}

In this work, we have clarified general structure of $\cN$-fold SUSY systems
by considering dimensional analysis and introducing the equivalent classes of
linear differential operators associated with them. We have then shown that
the latter general consideration is in fact effective in constructing the most
general $\cN$-fold SUSY systems and their integral constants for $\cN=2$, $3$,
and $4$. Application to systems for $\cN>4$ would be straightforward and
the problems would get more transparent even though still remain highly
complicated.
Finally, some remarks on the future issues are in order.\\

\noindent
1. Although we have only considered ordinary one-dimensional Schr\"{o}dinger
operators, generalization to other operators would be possible. In physical
applications, one of the interesting extensions is to a quantum system with
position-dependent mass for which $\cN$-fold SUSY was successfully
formulated in Ref.~\cite{Ta06a}. In the latter case, there is an additional freedom
of mass function and it is particularly interesting to see how its existence would
force us to modify or generalize the general considerations made in
Section~\ref{sec:gcon}.\\

\noindent
2. In this work, we have restricted the dimension-preserving transformations
(\ref{eq:2ftf}), (\ref{eq:3ftf}), and (\ref{eq:4ftf}) to polynomial type, namely,
transformations which are polynomials in $w_{k}^{[\cN](m)}$ ($m=0,1,2,\ldots$).
On the other hand, the obtained results such as (\ref{eq:2fu0}), (\ref{eq:3fu0}),
and (\ref{eq:4fu1}) indicate that the relations among $w_{k}^{[\cN](m)}$ are in
general expressed by rational functions. Hence, we may be able to reduce
further the complexity of the conditions for $\cN$-fold SUSY by extending
the type of transformations from polynomial to rational function. However,
the number of admissible forms of rational functions which preserve physical
dimension would drastically increase. For instance, any sum of rational
functions of the form $f_{n}^{[\cN]}[w]/g_{n}^{[\cN]}[w]$ ($n\in\bbZ$)
\begin{align*}
w_{k}^{[\cN]}=u_{k}^{[\cN]}+\sum_{n\in\bbZ}\frac{f_{n}^{[\cN]}[w]}{g_{n}^{[\cN]}[w]}
 \ [\rmL^{k-\cN}],
\end{align*}
where $f_{n}^{[\cN]}[w]$ and $g_{n}^{[\cN]}[w]$ are polynomials in $w_{k}^{[\cN](m)}$
having the physical dimensions $[\rmL^{n+k-\cN}]$ and $[\rmL^{n}]$, respectively,
can serve as a (part of) transformation which preserves the physical dimension 
$[\rmL^{k-\cN}]$ of $w_{k}^{[\cN]}$. As a result, we may need additional guidelines
to restrict the forms of transformations to make an efficient analysis. We are
curious to know how to get such guidelines systematically for the purpose.\\

\noindent
3. As was pointed out in Section~\ref{sec:gcon}, an $\cN$-fold SUSY system is
in general only weakly quasi-solvable but is not quasi-solvable in the strong
sense and thus does not necessarily admit analytic local solutions in closed form.
In the case of $\cN=2$, it was proved in Ref.~\cite{GT06} that type A $2$-fold
SUSY is a necessary and sufficient condition for a one-dimensional quantum
mechanical system to have quasi-solvability in the strong sense with two
independent analytic local solutions. So, it is interesting to see what kind of
condition is necessary and sufficient for a quantum system to admit three or
four independent analytic local solutions. We will report on the latter subjects
in subsequent publications.



\appendix

\section{Formulas}

In this Appendix, we present the lengthy formulas appeared in $4$-fold SUSY
in Section~\ref{sec:4f}.\\

\noindent
The third integral:
\begin{align}
\lefteqn{
1024J_{3}[w]=
 2w''_{3}w''''_{3}-(w'''_{3})^{2}+8w''''_{3}u_{1}+4w_{3}w''''_{3}u_{2}
 -4w'_{3}w'''_{3}u_{2}+6(w''_{3})^{2}u_{2}
}\notag\hspace{10pt}\\
&\;-2w_{3}w'''_{3}u'_{2}+6w'_{3}w''_{3}u'_{2}+2w_{3}w''_{3}u''_{2}+32w''_{3}u_{2}u_{1}
 +24w'_{3}u'_{2}u_{1}+8w_{3}u''_{2}u_{1}\notag\\
&\;+32u_{2}(u_{1})^{2}-4(w_{3})^{3}w''''_{3}+12(w_{3})^{2}w'_{3}w'''_{3}
 -16(w_{3})^{2}(w''_{3})^{2}-24w_{3}(w'_{3})^{2}w''_{3}\notag\\
&\;+(w'_{3})^{4}-80(w_{3})^{2}w''_{3}u_{1}-80w_{3}(w'_{3})^{2}u_{1}
 +16w_{3}w''_{3}(u_{2})^{2}-4(w'_{3})^{2}(u_{2})^{2}\notag\\
&\;+8w_{3}w'_{3}u_{2}u'_{2}+4(w_{3})^{2}u_{2}u''_{2}-(w_{3})^{2}(u'_{2})^{2}
 +32w_{3}(u_{2})^{2}u_{1}-56(w_{3})^{3}w''_{3}u_{2}\notag\\
&\;-20(w_{3})^{2}(w'_{3})^{2}u_{2}-4(w_{3})^{4}u''_{2}-64(w_{3})^{3}u_{2}u_{1}
 +8(w_{3})^{2}(u_{2})^{3}+40(w_{3})^{5}w''_{3}\notag\\
&\;+10(w_{3})^{4}(w'_{3})^{2}+48(w_{3})^{5}u_{1}-28(w_{3})^{4}(u_{2})^{2}
 +36(w_{3})^{6}u_{2}-15(w_{3})^{8}+32C_{1}(w'_{3})^{2}\notag\\
&\;+256C_{1}w_{3}u_{1}+64C_{1}(w_{3})^{2}u_{2}-32C_{1}(w_{3})^{4}+256(C_{1})^{2}
 =1024C_{3}\ [\rmL^{-8}].
\label{eq:4fC3}
\end{align}
$P_{4}^{\pm}$ and $V^{\pm}$ in terms of $w_{3}$ and $u_{2}$:
\begin{subequations}
\label{eqs:4fPV'}
\begin{align}
P_{4}^{\pm}\stackrel{4}{\sim}&\;\del^{4}\mp w_{3}\del^{3}+\left[u_{2}
 -\frac{3}{2}(w_{3})^{2}\mp\frac{3}{2}w'_{3}\right]\del^{2}\mp\biggl[\frac{3}{4}w''_{3}
 -\frac{w'''''_{3}}{4u'_{2}}-\frac{w'''_{3}u_{2}}{u'_{2}}-\frac{w'_{3}u''_{2}}{u'_{2}}
 -\frac{w_{3}u'''_{2}}{4u'_{2}}\notag\\
&\;+\frac{5(w_{3})^{2}w'''_{3}}{2u'_{2}}+\frac{10w_{3}w'_{3}
 w''_{3}}{u'_{2}}+\frac{5(w'_{3})^{3}}{2u'_{2}}-\frac{w'_{3}(u_{2})^{2}}{u'_{2}}
 +\frac{6(w_{3})^{2}w'_{3}u_{2}}{u'_{2}}-\frac{15(w_{3})^{4}w'_{3}}{2u'_{2}}\notag\\
&\;-\frac{8C_{1}w'_{3}}{u'_{2}}\mp (u'_{2}-3w_{3}w'_{3})\biggr]\del+\frac{1}{2}u''_{2}
 -\frac{5}{2}w_{3}w''_{3}-\frac{11}{8}(w'_{3})^{2}+\frac{1}{4}(u_{2})^{2}
 -\frac{3}{2}(w_{3})^{2}u_{2}\notag\\
&\;+\frac{15}{8}(w_{3})^{4}-\frac{w_{3}w'''''_{3}}{4u'_{2}}-\frac{w_{3}w'''_{3}u_{2}}{u'_{2}}
 -\frac{w_{3}w'_{3}u''_{2}}{u'_{2}}-\frac{(w_{3})^{2}u'''_{2}}{4u'_{2}}
 +\frac{5(w_{3})^{3}w'''_{3}}{2u'_{2}}\notag\\
&\;+\frac{10(w_{3})^{2}w'_{3}w''_{3}}{u'_{2}}+\frac{5w_{3}(w'_{3})^{3}}{2u'_{2}}
 -\frac{w_{3}w'_{3}(u_{2})^{2}}{u'_{2}}+\frac{6(w_{3})^{3}w'_{3}u_{2}}{u'_{2}}
 -\frac{15(w_{3})^{5}w'_{3}}{2u'_{2}}\notag\\
&\;-\frac{8C_{1}w_{3}w'_{3}}{u'_{2}}+2C_{1}\mp\left[\frac{7}{8}w'''_{3}
 +\frac{1}{2}w'_{3}u_{2}+\frac{3}{8}w_{3}u'_{2}-\frac{9}{4}(w_{3})^{2}w'_{3}\right],
\label{eq:P4+-'}\\
V^{\pm}=&\;-\frac{1}{4}u_{2}+\frac{1}{2}(w_{3})^{2}\pm\frac{1}{2}w'_{3}-C_{0}.
\label{eq:V4'}
\end{align}
\end{subequations}
The functions $f_{i}^{\pm}[w]$ in (\ref{eq:P4-+}) and (\ref{eq:P4+-}):
\begin{align}
\begin{split}
f_{4}^{+}=&\;-4I_{2},\qquad f_{3}^{+}=4I_{1}-4(3\del+w_{3})I_{2},\\
f_{2}^{+}=&\;-4I_{0}+(8\del+3w_{3})I_{1}-\frac{1}{2}[28\del^{2}+15w_{3}\del-9w'_{3}
 +4w_{2}+(w_{3})^{2}]I_{2},\\
f_{1}^{+}=&\;-2(2\del+w_{3})I_{0}+2(3\del^{2}+2w_{3}\del-2w'_{3}+w_{2})I_{1}\\
&\;-2[4\del^{3}+3w_{3}\del^{2}-2(2w'_{3}-w_{2})\del-4w''_{3}+w_{1}-2w_{3}w'_{3}]I_{2},\\
f_{0}^{+}=&\;-2(\del^{2}+w_{3}\del-2w'_{3}+w_{2})I_{0}+[2\del^{3}+2w_{3}\del^{2}
 -2(2w'_{3}-w_{2})\del-4w''_{3}+w_{1}\\
&\;-2w_{3}w'_{3}]I_{1}-\frac{1}{16}[32\del^{4}+32w_{3}\del^{3}-32(2w'_{3}-w_{2})\del^{2}
 -4(31w''_{3}-2w'_{2}-6w_{1}\\
&\;+14w_{3}w'_{3})\del-60w'''_{3}+8w''_{2}+16w_{0}-68w_{3}w''_{3}+7(w'_{3})^{2}-4w'_{3}w_{2}
 +8w_{3}w'_{2}\\
&\;-4(w_{2})^{2}-22(w_{3})^{2}w'_{3}+4(w_{3})^{2}w_{2}-(w_{3})^{4}]I_{2},
\end{split}
\label{eq:f+}
\end{align}
and
\begin{align}
\begin{split}
f_{4}^{-}=&\;4I_{2},\qquad f_{3}^{-}=4I_{1}+4(\del-w_{3})I_{2},\\
f_{2}^{-}=&\;4I_{0}+(4\del-3w_{3})I_{1}+\frac{1}{2}[4\del^{2}-3w_{3}\del-3w'_{3}
 +4w_{2}-(w_{3})^{2}]I_{2},\\
f_{1}^{-}=&\;2(2\del-w_{3})I_{0}+2(\del^{2}-w_{3}\del-w'_{3}+w_{2})I_{1}
 +2(w''_{3}+2w'_{2}-w_{1}-2w_{3}w'_{3})I_{2},\\
f_{0}^{-}=&\;2(\del^{2}-w_{3}\del-w'_{3}+w_{2})I_{0}+(w''_{3}+2w'_{2}-w_{1}-2w_{3}w'_{3})
 I_{1}-\frac{1}{16}[4(w''_{3}+2w'_{2}\\
&\;-2w_{1}-2w_{3}w'_{3})\del-12w'''_{3}-24w''_{2}+16w_{0}+28w_{3}w''_{3}+23(w'_{3})^{2}
 -4w'_{3}w_{2}\\
&\;+8w_{3}w'_{2}-4(w_{2})^{2}-6(w_{3})^{2}w'_{3}+4(w_{3})^{2}w_{2}-(w_{3})^{4}]I_{2},
\end{split}
\label{eq:f-}
\end{align}
where $I_{2}$, $I_{1}$, and $I_{0}$ are given by (\ref{eq:4fc3'}), (\ref{eq:4fc4'}), and
(\ref{eq:4fc5'}), respectively.


\bibliography{refsels}

\begin{thebibliography}{10}
\expandafter\ifx\csname url\endcsname\relax
  \def\url#1{{\tt #1}}\fi
\expandafter\ifx\csname urlprefix\endcsname\relax\def\urlprefix{URL }\fi
\providecommand{\eprint}[2][]{\url{#2}}

\bibitem{AIS93}
A.~A. Andrianov, M.~V. Ioffe, and V.~P. Spiridonov, Phys. Lett. A 174 (1993)
  273.
\newblock \eprint{arXiv:hep-th/9303005}.

\bibitem{AST01b}
H.~Aoyama, M.~Sato, and T.~Tanaka, Nucl. Phys. B 619 (2001) 105.
\newblock \eprint{arXiv:quant-ph/0106037}.

\bibitem{AS03}
A.~A. Andrianov and A.~V. Sokolov, Nucl. Phys. B 660 (2003) 25.
\newblock \eprint{arXiv:hep-th/0301062}.

\bibitem{Ta03a}
T.~Tanaka, Nucl. Phys. B 662 (2003) 413.
\newblock \eprint{arXiv:hep-th/0212276}.

\bibitem{Ta09}
T.~Tanaka, In Morris~B. Levy, ed., Mathematical Physics Research Developments
  (Nova Science Publishers, Inc., New York, 2009), chapter~18. pp. 621--679.

\bibitem{GT05}
A.~Gonz{\'a}lez-L{\'o}pez and T.~Tanaka, J. Phys. A: Math. Gen. 38 (2005) 5133.
\newblock \eprint{arXiv:hep-th/0405079}.

\bibitem{AST01a}
H.~Aoyama, M.~Sato, and T.~Tanaka, Phys. Lett. B 503 (2001) 423.
\newblock \eprint{arXiv:quant-ph/0012065}.

\bibitem{GT04}
A.~Gonz{\'a}lez-L{\'o}pez and T.~Tanaka, Phys. Lett. B 586 (2004) 117.
\newblock \eprint{arXiv:hep-th/0307094}.

\bibitem{Ta10a}
T.~Tanaka, J. Math. Phys. 51 (2010) 032101.
\newblock \eprint{arXiv:0910.0328 [math-ph]}.

\bibitem{Cr55a}
M.~M. Crum, Quart. J. Math. 6 (1955) 121.

\bibitem{AICD95}
A.~A. Andrianov, M.~V. Ioffe, F.~Cannata, and J.~P. Dedonder, Int. J. Mod.
  Phys. A 10 (1995) 2683.
\newblock \eprint{arXiv:hep-th/9404061}.

\bibitem{AIN95b}
A.~A. Andrianov, M.~V. Ioffe, and D.~N. Nishnianidze, Phys. Lett. A 201 (1995)
  103.
\newblock \eprint{arXiv:hep-th/9404120}.

\bibitem{IN04}
M.~V. Ioffe and D.~N. Nishnianidze, Phys. Lett. A 327 (2004) 425.
\newblock \eprint{arXiv:hep-th/0404078}.

\bibitem{AKOSW99}
H.~Aoyama, H.~Kikuchi, I.~Okouchi, M.~Sato, and S.~Wada, Nucl. Phys. B 553
  (1999) 644.
\newblock \eprint{arXiv:hep-th/9808034}.

\bibitem{ST02}
M.~Sato and T.~Tanaka, J. Math. Phys. 43 (2002) 3484.
\newblock \eprint{arXiv:hep-th/0109179}.

\bibitem{Ta06a}
T.~Tanaka, J. Phys. A: Math. Gen. 39 (2006) 219.
\newblock \eprint{arXiv:quant-ph/0509132}.

\bibitem{GT06}
A.~Gonz{\'a}lez-L{\'o}pez and T.~Tanaka, J. Phys. A: Math. Gen. 39 (2006) 3715.
\newblock \eprint{arXiv:quant-ph/0602177}.

\end{thebibliography}
\bibliographystyle{npb}



\end{document}